\documentclass[twocolumn,aps,showpacs,prb]{revtex4-1}
\usepackage{graphicx}
\usepackage{subfigure}
\usepackage{epstopdf}
\usepackage{amsmath}
\usepackage{multirow}
\usepackage{tabularx}
\usepackage{color,ulem}
\begin{document}
\title{First-principles study of the electronic structure of CaKRu$_4$P$_4$}
\author{Xinlei Zhao$^{1}$}
\author{Fengjie Ma$^{1}$}\email{fengjie.ma@bnu.edu.cn}
\author{Zhong-Yi Lu$^{2}$}

\date{\today}

\affiliation{$^{1}$The Center for Advanced Quantum Studies and Department of Physics, Beijing Normal University, 100875 Beijing, China}
\affiliation{$^{2}$Department of Physics, Renmin University of China, Beijing 100872, China}

\begin{abstract}

The recent discovery and studies of 1144-phase compounds, e.g., CaKFe$_4$As$_4$, have attracted significant research interest. In this paper, based on the first-principles density functional calculations, we present a systematic study on the electronic structure of the recently synthetized 1144-type quaternary compound CaKRu$_4$P$_4$. We find that the Ru-based 1144-type compound possesses a different electronic structure from that of iron-based superconductors, even though they share very similar crystallographic structures. In CaKRu$_4$P$_4$, there is no hole-type carrier if spin-orbit interaction is not considered. And a long-range magnetic order is absent in its ground state. With the application of pressure, the electronic structure of CaKRu$_4$P$_4$ becomes similar to those of the ternary 122-type compounds LaRu$_2$P$_2$ and LaRu$_2$As$_2$. CaKRu$_4$P$_4$ is very likely to be a phonon-mediated medium coupled BCS superconductor. Furthermore, type-I and type-II Dirac fermions can be created and regulated in this system with pressure. The quaternary compound CaKRu$_4$P$_4$ therefore has a potential to be an attractive platform for the study of topological physics and superconductivity.

\end{abstract}

\maketitle

\section{Introduction}

The discovery of iron-based superconductors has generated enormous interest over the past decade \cite{kamihara2008iron-based, PhysRevLett.101.107006, wang2008the, hsu2008superconductivity, Iyo2016New, PhysRevB.82.180520, Bao_2011, RN1333, RN768, RN743, Paglione2010High, dai2012magnetism, RevModPhys.84.1383, RN994, Chen2014Iron, RevModPhys.87.855, RN764, kordyuk2012iron-based}. Besides the high-temperature superconductivity, these materials also host the possibility of realizing topological quantum computing \cite{10.1093/nsr/nwy142, RN1321}. Their nontrivial bulk topology and high-temperature superconductivity can interact to give rise to topological superconductivity capable of supporting the emergence of zero-energy-gap quasiparticle excitations, named Majorana fermions, within the bulk superconducting gap. The Majorana fermions satisfy non-Abelian statistics, which make them a prime candidate for realizing the topological quantum computing and becoming one of the most active fields in condensed matter physics \cite{Zhang182, Wang333, Zhu189, RN1322}.

Recently, a new family of the 1144-type quaternary iron arsenides, AeAFe$_4$As$_4$ (Ae=Ca, Sr, Ba, La, Eu; A=Na, K, Rb, Cs), was reported to show superconductivity with transition temperatures $T_{c}$ $\sim$ 24-37 K, which has attracted great interest \cite{Iyo2016New, PhysRevLett.126.157001, NEJADSATTARI2020109137, PhysRevB.93.214503, LIU20161213, RN1331, PhysRevB.94.064501, PhysRevMaterials.1.013401, PhysRevB.95.140505, PhysRevB.95.100502, PhysRevLett.117.277001, RN1329, RN1330, PhysRevResearch.2.022018, PhysRevB.97.094105}. It shares a common crystalline feature with the 122-type iron-based superconductors, and can be regarded as a hybrid phase between AeFe$_2$As$_2$ and AFe$_2$As$_2$. The Ae and A layers alternately intersect in the $c$-axis direction of AeAFe$_4$As$_4$, and occupy crystal nonequivalent sites, leading to a change in the space group from $I4$/$mmm$ to $P4$/$mmm$ \cite{PhysRevLett.101.107006,Iyo2016New, PhysRevB.84.054502}. It has been suggested that the surface of 1144-type crystals could be a promising platform for testing the Majorana fermions, and hence for realization of topological quantum computing \cite{RN1322}. The search for more materials with structures analogous to 1144-type iron-based superconductors is therefore significantly important, since it may trigger further discovery of new topological superconductors and novel emergent properties.

\begin{figure}
\centering
\includegraphics[width=8.5cm]{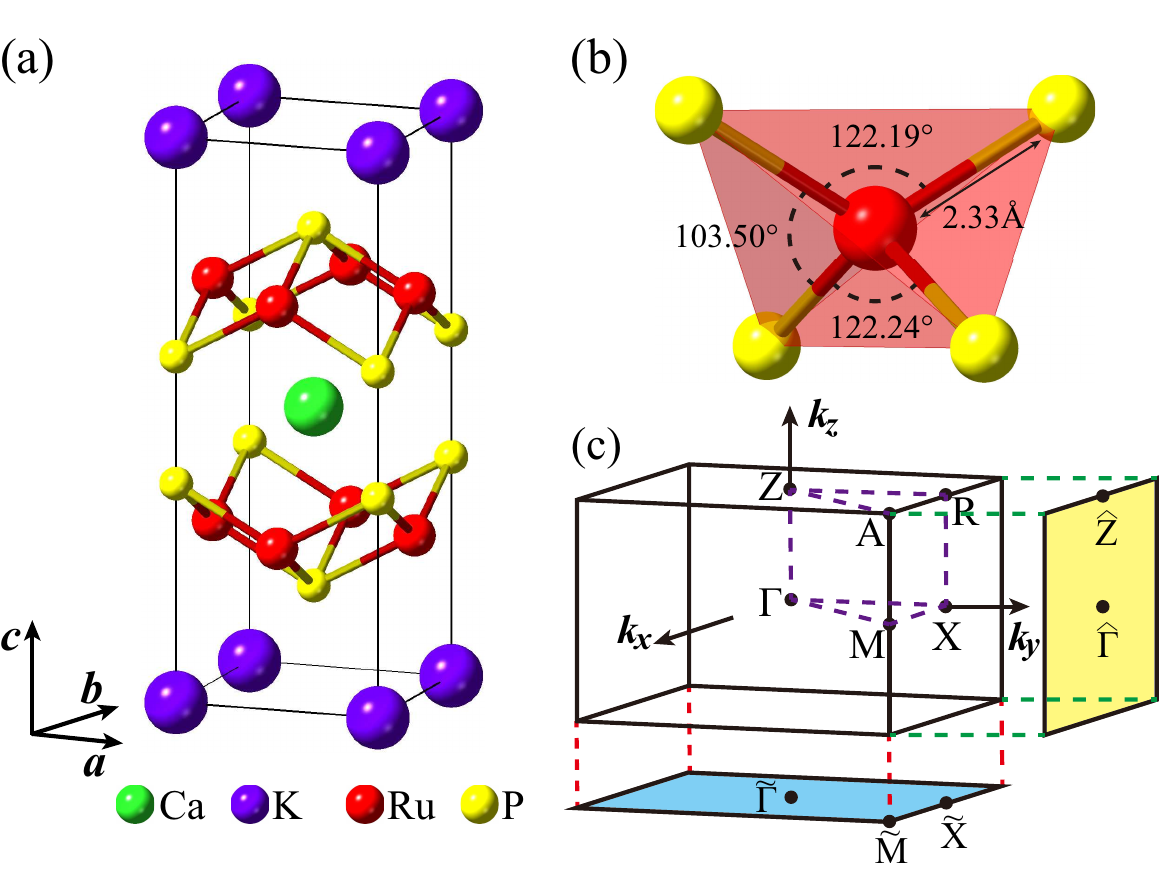}
\caption{\label{crystal} (a) Tetragonal unit cell of CaKRu$_4$P$_4$. The principal axes $a$, $b$, and $c$ are indicated. (b) Diagram of bond angles between P-Ru-P and bond length between Ru-P. (c) The bulk Brillouin zone (BZ) with the high symmetry points labeled. The projected surface BZ of (001) and (010) planes are depicted by blue and yellow planes, respectively.}
\end{figure}

Since the element Ru locates in the same group as the element Fe in the Periodic Table, the similar physical properties are expected in the Ru-based and Fe-based compounds. There have been a number of studies focused on the Ru-related superconductors to date \cite{RN1320, JEITSCHKO198793, GUO2016921}. For example, the layered perovskite Sr$_2$RuO$_4$ is a well-known unconventional superconductor with $T_c$ of 0.93 K \cite{RN1320}. The ternary 122-type compounds LaRu$_2$P$_2$ and LaRu$_2$As$_2$ were also reported to show superconductivity with $T_c$ of 4.1 K and 7.8 K, respectively \cite{JEITSCHKO198793, GUO2016921}. Recently, a new high-purity 1144-type Ru-based material, CaKRu$_4$P$_4$, was successfully synthesized in experiment \cite{song2021hetero}. The study of such a 1144-type Ru-based material and comparison it with the discovered 122-type or other 1144-type systems is very meaningful for achieving topological superconductivity.

In this paper, we have systemically studied the electronic and structural properties of CaKRu$_4$P$_4$ based on the first-principles density functional theory. In comparison with the iron-based superconductors, the Ru-based 1144-type compound has different electronic structure, although they share very similar crystallographic structures. Under ambient condition, there are only electron-type carriers in CaKRu$_4$P$_4$ when spin-orbit coupling (SOC) is not considered. More importantly, a long-range magnetic order is absent in the material. Because of the similar electronic, structural, and nonmagnetic properties to those of the ternary 122-type Ru-based compounds LaRu$_2$P$_2$ and LaRu$_2$As$_2$, CaKRu$_4$P$_4$ is very likely to be a phonon-mediated medium coupled BCS superconductor under pressure \cite{JEITSCHKO198793, GUO2016921, PhysRevLett.108.257005, RAHAMAN20172827, PhysRevB.84.224507, Hadi_2017}. Furthermore, topological states, including both type-I and type-II Dirac fermions, can be created and regulated with pressure in this material. The quaternary compound CaKRu$_4$P$_4$ therefore has the potential to be an attractive platform for the study of topological physics, superconductivity, and topological superconductivity.

\section{Computational details}

In our calculations, the plane-wave basis method and QUANTUM ESPRESSO software package were used \cite{QE2009, QE2017}. We adopted the generalized gradient approximation of the Perdew-Burke-Ernzerhof formula for the exchange-correlation potentials in the electronic structure calculations \cite{perdew1996generalized}. The ultrasoft pseudopotentials were employed to model the electron-ion interactions \cite{vanderbilt1990soft}. After the full convergence test, the kinetic energy cutoff for wave functions and charge density were chosen to be 816 and 6530 eV, respectively. The Marzari-Vanderbilt broadening technique was used \cite{marzari1999thermal}. For the density of states (DOS) calculations, a mesh of \mbox{24$\times$24$\times$12} $k$ points and the tetrahedra method were used \cite{PhysRevB.49.16223}. All of the lattice parameters and internal atomic positions were fully optimized. 
For the calculations under pressure, an external hydrostatic pressure was applied on the simulation cell, whose geometries were fully relaxed by minimizing the enthalpy. The topological surface states were studied using tight-binding methods by the combination of Wannier90 \cite{Pizzi2020} and WannierTools \cite{WU2017} software packages.

\section{Results and discussion}

\begin{figure}
\centering
\includegraphics[width=8.5cm]{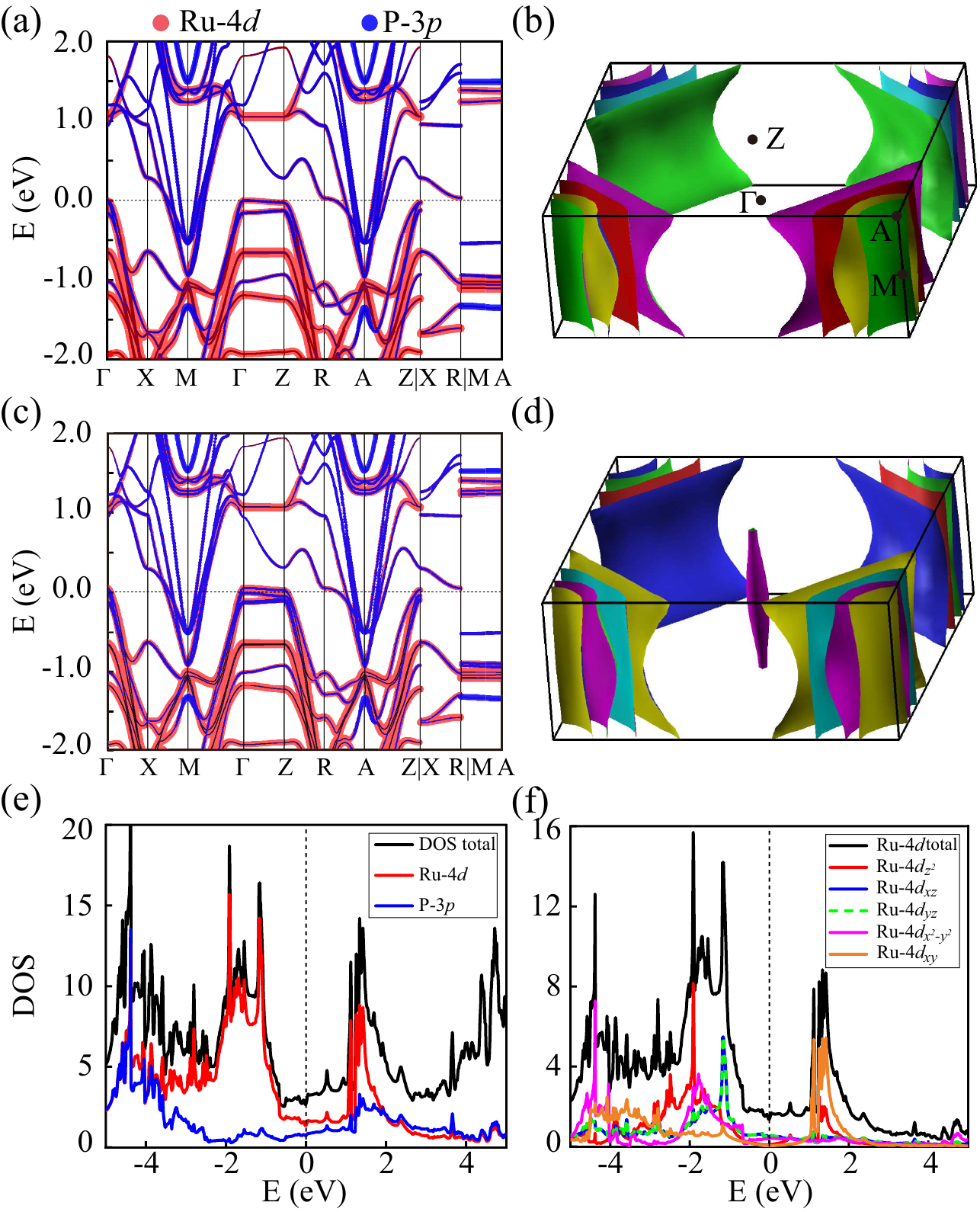}
\caption{\label{bands} (a) Band structure of CaKRu$_4$P$_4$ without SOC. The red and blue colored lines represent the projections of Ru-4$d$ and P-3$p$ orbitals, respectively, whose linewidths indicate the weight of each component. (b) The Fermi surface of CaKRu$_4$P$_4$ without SOC. When SOC is included, the corresponding band structure and Fermi surface of CaKRu$_4$P$_4$ are given in (c) and (d), respectively. (e) The total and orbital-resolved partial density of states (DOS) of CaKRu$_4$P$_4$. (f) The projected DOS of Ru-4d orbitals. The Fermi energy is set to zero.}
\end{figure}

CaKRu$_4$P$_4$ has a tetragonal layered crystal structure with $P4$/$mmm$ (No. 123) space group, as shown in Fig. \ref{crystal}(a). Each Ru atom bonds with four surrounding P atoms to form a tetrahedron corner-shared connected with each other, similar to the FeAs or FeSe structure in the iron-based superconductors \cite{PhysRevLett.101.107006,hsu2008superconductivity}. The layers of RuP tetrahedrons are alternately separated by Ca or K ions along the $c$ axis. If the positions of Ca and K are occupied by the same atoms, the crystal structure of CaKRu$_4$P$_4$ is the same as that of 122-type iron-based superconductors. However, different from the most of iron-based superconductors \cite{RN1334, Dong_2008, PhysRevLett.102.177003, PhysRevLett.102.247001, PhysRevB.79.054503, PhysRevB.83.233205, RN631, RN1142}, our calculations show that the ground state of CaKRu$_4$P$_4$ is nonmagnetic, although there is $4d$ transition-metal elementary composition in the material.  In our calculations, the optimized lattice constants are about $a$=$b$=4.08 $\AA$ and $c$=11.03 $\AA$, which are in good agreement with the experimental results \cite{song2021hetero}. As shown in Fig. \ref{crystal}(b), the bond length between Ru and P atoms is 2.33 $\AA$ and the bond angles of P-Ru-P are 122.19$^{\circ}$, 103.50$^{\circ}$, and 122.24$^{\circ}$. The RuP tetrahedrons are distorted from the ideal bond angle value of 109.47$^{\circ}$, and are also significantly different from the numbers (112.1$^{\circ}$ and 107.6$^{\circ}$) of 1144-type iron-based superconductor CaKFe$_4$As$_4$ \cite{Iyo2016New}. It was found that in iron-based superconductors the magnetic moments of the transition-metal ions are sensitive to the bond angles between the cations and anions of the tetrahedrons \cite{angle}. This could be one reason why the ground state of CaKRu$_4$P$_4$ is nonmagnetic. To further simplify the corresponding physics, here we focus on the RuP tetrahedron. With the changes of bond angles, 
the tetrahedron is distorted toward square planar geometry. The degeneracy of $t_{2g}$/$e_g$ orbitals under tetrahedral crystal field breaks, in which the $d_{xz/yz}$ of the high energetic $t_{2g}$ orbitals goes lower while the $d_{xy}$ of the low energetic $e_g$ orbital is pushed upward (Ru-Ru bond is chosen along the diagonal directions). These three orbitals become close to each other, and could even be reversed in the order of energy level. In addition, Hund's coupling is weaker in the $4d$ transition metal compounds than in the $3d$ ones. Both of these effects prefer to reduce the magnetic moments of $d^6$ Ru atoms, leading to a nonmagnetic ground state.

The band structure of CaKRu$_4$P$_4$ under standard ambient condition is shown in Fig. \ref{bands}(a), which exhibits a metallic nature with predominantly Ru-4$d$ and P-3$p$ orbitals around the Fermi level. The energy dispersions of bands along the $k_z$ axis ($\Gamma$-$Z$ or $M$-$A$) near the Fermi level are very small in comparison with the in-plane ones, resulting in two-dimensional-like Fermi surfaces consisting of four electron pockets at the corners of the Brillouin zone (BZ), as shown in Fig. \ref{bands}(b). This is a little different from most iron-based superconductors, in which there are both electron-type and hole-type carriers in the parent compounds. Note that there are two doubly-degenerated valence bands located right below the Fermi level along $\Gamma$-$Z$. Since the Ru element has significant spin-orbit effect that could induce band splittings, one split band is pushed upward crossing the Fermi level when SOC is considered, as shown in Fig. \ref{bands} (c). One more hole-type Fermi surface sheet is thus formed centered around $\Gamma$-$Z$, as shown in Fig. \ref{bands} (d). From the volumes enclosed by these Fermi surfaces, we find that the electron and hole carrier density are about 5.53 $\times 10^{21}/cm^3$ and 2.02 $\times 10^{19}/cm^3$, respectively. Figures \ref{bands}(e) and \ref{bands}(f) give the total and orbital-resolved partial DOS of CaKRu$_4$P$_4$. The states around the Fermi level contribute mainly by the Ru-4$d$ and P-3$p$ orbitals, which is consistent with the orbital projections of the band structure [Fig. \ref{bands}(a)]. At the Fermi level the DOS is about 2.86 states/eV, which yields the electronic specific heat coefficient $\gamma$ = 6.74 $mJ/(K^2 \cdot mol)$ and Pauli paramagnetic susceptibility $\chi_p$ = 1.16 $\times 10^{-9}m^3/mol$. Because of the weak tetrahedral crystal field imposed on the Ru atom by the surrounding four P atoms, the five 4$d$ orbitals of Ru are nearly uniformly distributed and partially filled, different from that in transition metal oxides.


\begin{figure}
\centering
\includegraphics[width=7.0cm]{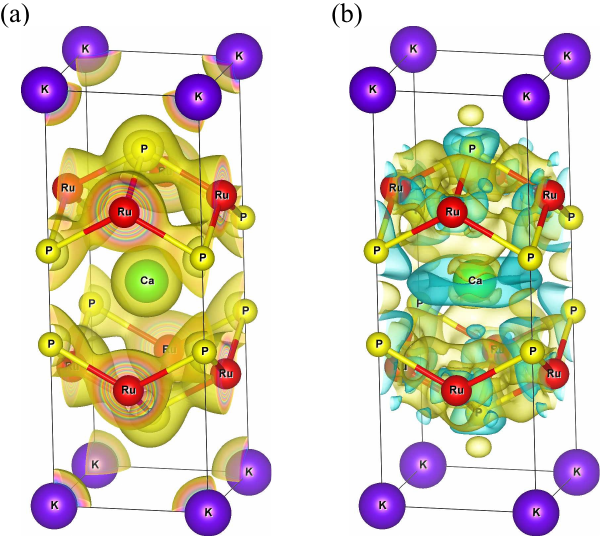}
\caption{\label{charge} (a) Charge density distribution ($e$/$bohr^3$) of CaKRu$_4$P$_4$. The isosurface value is 0.08 $e/\AA^3$. (b) The differential charge density distribution ($e$/$bohr^3$) of CaKRu$_4$P$_4$. The yellow and cyan colored regions indicate gain or loss of electrons, respectively. The isosurface value is 0.004 $e/\AA^3$.}
\end{figure}

Figures \ref{charge}(a) and \ref{charge}(b) show the charge and differential charge density distributions of CaKRu$_4$P$_4$, respectively. Large overlaps between the electronic clouds of Ru and P atoms are found, suggesting that there are strong hybridizations between Ru and P orbitals along the Ru-P bonds. Instead, the direct overlap between two neighboring Ru atoms is negligibly small. Moreover, from the differential charge density distribution, we find that the differential charges are more likely accumulated around P atoms and decreased around Ru atoms, which is consistent with the electronegativity of each element. The Ca layers in the material serve as a charge reservoir, while the K atoms contribute very little to the differential charge density. Due to the balance of charge valences, the Ru atoms are fractional valenced in the compound.

\begin{figure}
\centering
\includegraphics[width=8.5cm]{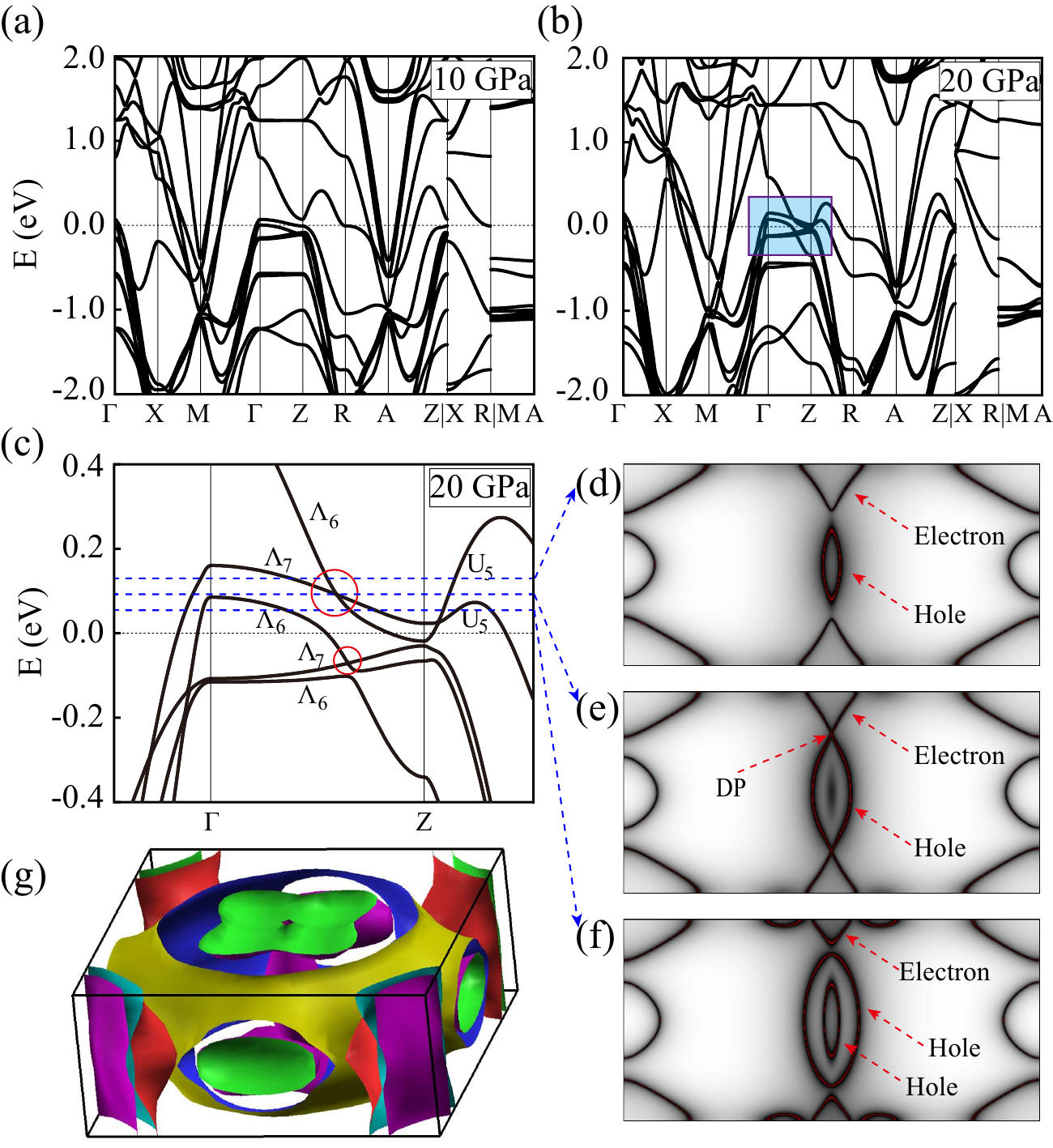}
\caption{\label{pressure} Calculated band structures of CaKRu$_4$P$_4$ under hydrostatic pressures of (a) 10 GPa and (b) 20 GPa with SOC. (c) The enlargement of band structure in the box of subfigure (b) with irreducible representations labeled. The Dirac points along the $\Gamma$-$Z$ path are marked with red circles. (d)-(f) The contours of electron and hole pockets around the Dirac point (DP) in the ${k_x}$-${k_z}$ plane at $E_{DP}$+40 meV, $E_{DP}$, and $E_{DP}$-40 meV, respectively. $E_{DP}$ is the energy at which the Dirac point locates. (g) The Fermi surface of CaKRu$_4$P$_4$ at 20 GPa.
}
\end{figure}

We have also studied the evaluation of the electronic structure of CaKRu$_4$P$_4$ under pressure, which is one of the fundamental ways used widely to perturb the properties of iron-based superconductors \cite{PhysRevB.78.184517,PhysRevB.88.045117,PhysRevB.79.060510,Zhang182, Wang333, Zhu189}. The emergence of a pressure induced topological state is found (SOC is always included). When a hydrostatic pressure is applied, there are stronger interlayer interactions between the Ca/K layers and RuP layers because of the compression along the $c$ axis. The valence bands along $\Gamma$-$Z$ shift up gradually and eventually intersect with the conduction bands between 10 and 20 GPa, as shown in Figs. \ref{pressure}(a) and \ref{pressure}(b). Two crossings are formed near the Fermi level. Since these bands are all doubly degenerated due to the protection of time-reversal and space-inversion symmetries, the crossing points are fourfold degenerated. As labeled in Fig. \ref{pressure}(c), their irreducible representations along $\Gamma$-$Z$ near the Fermi level are $\Lambda_6$ and $\Lambda_7$, belonging to the $D_{4h}$ double group. Therefore, these crossings are the Dirac points protected by the $D_{4h}$ symmetry. The other intersections along $X$-$M$-$\Gamma$ and $R$-$A$-$Z$ are opened, since the bands in the paths belong to the same irreducible representation of the $C_{2v}$ double group. Note that multiple types of topological fermions can be found in this material. As marked by the red circles in Fig. \ref{pressure}(c), the Dirac cone above the Fermi level is tilted strongly along $\Gamma$-$Z$, which is a typical feature of the type-II Dirac fermion \cite{Yan2017}. Two open electronlike and holelike pockets touching at the Dirac point could be created if the Fermi level crosses the intersecting point. The evolution of hole and electron pockets with different chemical potential in the $k_x$-$k_z$ plane is shown in Figs. \ref{pressure}(d)-(f). They will only be touched at the energy of the type-II Dirac point ($E_{DP}$) and separated elsewhere. The Fermi surface composed of nontrivial electron and hole pockets may contribute to the formation of Cooper pairs and make the compound superconducting \cite{PhysRevB.95.094513}. By contrast, the other cone formed by the valence bands belongs to the type-I Dirac fermion, which possesses a pointlike Fermi surface geometry when the Dirac point lies at the Fermi level. Figure \ref{pressure}(g) gives the Fermi surface of CaKRu$_4$P$_4$ at 20 GPa, which is more three-dimensional-like compared with the one under a standard ambient condition. Due to the bands shift, there are two more hole-type Fermi surface sheets, one of which is centered around $\Gamma$-$Z$, while the other one forms a pocket around $X$. The electronic structure of CaKRu$_4$P$_4$ under pressure is very similar to that of 122-type Ru-based superconductors LaRu$_2$P$_2$ and LaRu$_2$As$_2$ \cite{PhysRevLett.108.257005, RAHAMAN20172827, PhysRevB.84.224507, Hadi_2017}. Therefore, CaKRu$_4$P$_4$ shares very similar crystal, electronic, and nonmagnetic properties with those of LaRu$_2$P$_2$ and LaRu$_2$As$_2$, which are closely related to the superconductivity of materials. Since LaRu$_2$P$_2$ and LaRu$_2$As$_2$ are both found to be phonon-mediated medium coupled BCS superconductors \cite{PhysRevLett.108.257005, RAHAMAN20172827, PhysRevB.84.224507, Hadi_2017}, likewise CaKRu$_4$P$_4$ is a BCS superconductor. As a comparison, the 1144-type and 122-type iron-based compounds CaKFe$_4$As$_4$, CaFe$_2$As$_2$, SrFe$_2$As$_2$, and BaFe$_2$As$_2$ all have very similar structural, electronic, and As-bridged antiferromagnetic properties, and all are unconventional high-Tc superconductors \cite{Iyo2016New, RevModPhys.87.855,RN994,Chen2014Iron, RevModPhys.84.1383}.

\begin{figure}
\centering
\includegraphics[width=8.5cm]{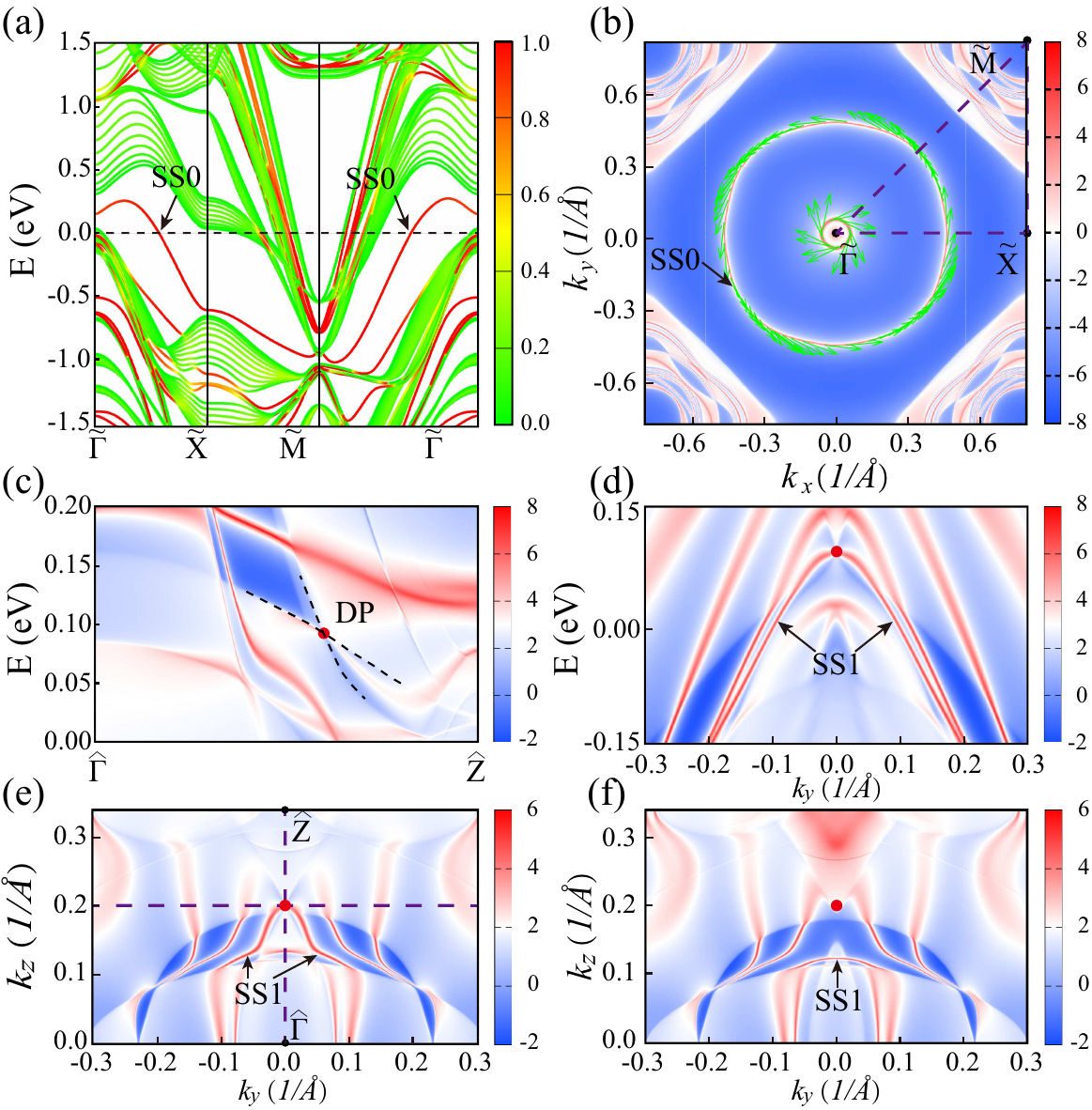}
\caption{\label{surface} (a) Band dispersion of CaKRu$_4$P$_4$ on the projected (001) surface along the high-symmetry lines (purple dashed lines) marked in the subfigure (b) under ambient pressure. Green and red colors represent the bulk and surface components, respectively. (b) The constant-energy surfaces at the Fermi level under ambient pressure. The surface states passing through the Fermi level, SS0, are indicated by the black arrow and their spin textures are represented by the green arrows. (c) The projected surface states onto the (010) surface along $\widehat{\Gamma}$-$\widehat{Z}$ [the vertical dashed line in the subfigure (e)] at 20 GPa. The black dashed lines are a guide to the eye illustrating the bulk bands' dispersion around the type-II Dirac point (red dot). (d) The projected surface states onto the (010) surface along the horizontal dashed line passing through the Dirac point in the subfigure (e). (e),(f) The constant-energy surfaces at $E_{DP}$ and $E_{DP}$+40 meV at 20 GPa, respectively. The surface states, SS1, are indicated by the black arrows.}
\end{figure}

The topological surface states (SSs) of CaKRu$_4$P$_4$ are studied with the WannierTools software package based on the Green$'$s function method using a tight-binding Hamiltonian constructed by the maximally localized Wannier functions \cite{WU2017, marzari2012maximally}. Figure \ref{surface}(a) gives the band dispersions projected onto the (001) surface under ambient pressure, following the high-symmetry $k$ paths marked by the purple dashed lines in Fig. \ref{surface}(b). There is a SS band (labeled SS0) crossing the Fermi level, which forms a loop with spin-momentum locked spin texture on the projected Fermi surface at the Fermi level, similar to the topological insulators \cite{RN1352, PhysRevLett.119.077702}. The SSs are robust until the local energy gap of bulk states around -1 eV at $M$ is closed around 5 GPa. Figures \ref{surface}(c) and \ref{surface}(d) illustrate the band dispersions projected onto the (010) surface at 20 GPa, along the $k$ paths marked by the vertical dashed and horizontal dashed lines in Fig. \ref{surface}(e), respectively. A tilting type-II Dirac cone can be found in Fig. \ref{surface}(c), as guided by the black dashed lines, while the type-I Dirac point is completely burried in the bulk continuum. Topological nontrivial SSs originating from the type-II Dirac point (labeled as SS1) are clearly observed in Fig. \ref{surface}(d). Figures \ref{surface}(e) and \ref{surface}(f) show the constant-energy surfaces at $E_{DP}$ and $E_{DP}$+40 meV, respectively. Since a Dirac point can be viewed as the superposition of two Weyl points, there will be double Fermi arcs, SS1, originating from the Dirac point. At $E_{DP}$, the two Fermi arc surface states connect the projections of Dirac points onto the surface Brillouin zone. Interestingly, by varying the chemical potential, we can see that the Fermi arcs can be completely deformed into Fermi contours, because of the fragilite nature of Fermi arcs in Dirac semimetals \cite{PhysRevB.99.161113, Le8311,PhysRevB.100.245151}.

\section{Conclusion}

In summary, we have studied the electronic structures of the quaternary Ru-based 1144-type compound CaKRu$_4$P$_4$ based on first-principles calculations. Different from the case of iron-based superconductors, there is no long-range magnetic ordering in this material. Its electronic, structural, and nonmagnetic properties are similar to those of 122-type Ru-based superconductors LaRu$_2$P$_2$ and LaRu$_2$As$_2$. CaKRu$_4$P$_4$ is therefore very likely to be a phonon-mediated medium coupled BCS superconductor. Moreover, both type-I and type-II Dirac fermions can be realized in the system when high pressure is applied. The combination of topological and possible superconducting properties provides this material a promising platform for the study of topological superconductors.

\section{Acknowledgments}

This work was supported by the National Natural Science Foundation of China under Grants No. 12074040, No.11934020, and No. 11674027. All calculations were performed at the high performance computing cluster of the Center for Advanced Quantum Studies and Department of Physics, Beijing Normal University.



\bibliography{reference}

\end{document}